\documentclass[letterpaper,twocolumn,english,prl]{revtex4}

\usepackage{amsmath,amssymb,amsthm}
\usepackage{hyperref}
\usepackage{amsthm}
\usepackage{graphicx}
\usepackage{subfigure}
\usepackage{color,bbm}
\usepackage{amsfonts}
\usepackage{mathrsfs}
\usepackage{pdfpages}

\begin{document}
\date{\today}
\title{\bf  Second law of thermodynamics  at stopping times   }

\author{{\normalsize{}Izaak Neri}}
\affiliation{\noindent  \textit{ Department of Mathematics, King’s College London, Strand, London, WC2R 2LS, UK}}
\begin{abstract}     
Events in mesoscopic systems often take place at first-passage times, as is for instance the case for a
colloidal particle that escapes a metastable state. An interesting question is how much work an external
agent has done on a particle when it escapes a metastable state. We develop a thermodynamic theory for
processes in mesoscopic systems that terminate at stopping times, which generalize first-passage times.
This theory implies a thermodynamic bound, reminiscent of the second law of thermodynamics, for the
work exerted by an external protocol on a mesoscopic system at a stopping time. As an illustration, we use
this law to bound the work required to stretch a polymer to a certain length or to let a particle escape from a
metastable state.
 \end{abstract}

\maketitle

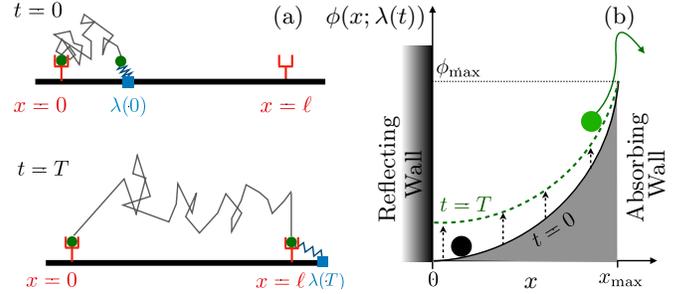
\begin{figure}[t]
\centering
{\includegraphics[width=0.5\textwidth]{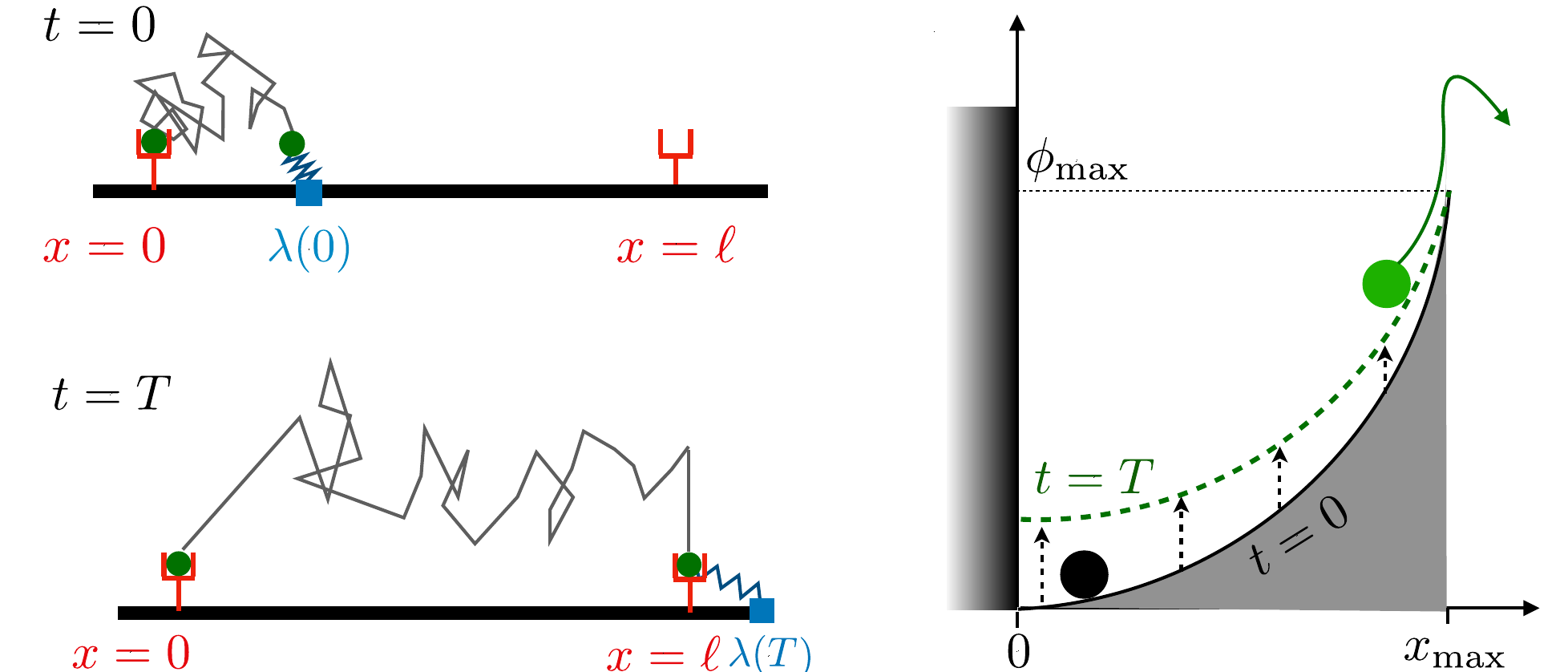}}
\put(-150,100){(a)}
\put(-25,100){(b)}
\put(-110,25){\rotatebox{90}{Reflecting}}
\put(-100,34){\rotatebox{90}{Wall}}
\put(-16,25){\rotatebox{90}{Absorbing}}
\put(-8,34){\rotatebox{90}{Wall}}
\put(-130,100){$\phi(x;\lambda(t))$}
\put(-55,0){$x$}
\caption{Stretching a  polymer to a certain length $\ell$ (Panel (a)) or  letting a particle escape from a metastable state  (Panel (b)).   Panel (a):     An external agent (blue square) is connected with a spring (blue zigzag line)  to one of the end points (green circles) of a polymer (grey zigzag line)  and stretches the polymer until it reaches a length $\ell$, after which the    polymer  end point is attached to an anchor point  (red object).    Panel (b): a colloidal particle (full circle) escapes from a metastable state  under the influence of an external protocol $\lambda(t)$ that changes the shape of the potential $\phi(x;\lambda)$.   } \label{fig1}
\end{figure}

\paragraph{Introduction.}    
How much work do we need to do on a mesoscopic system in order to let a certain event of interest happen?      For example,  how much work do we require  to stretch a polymer to a certain length  or to let a colloidal  particle  escape from a metastable state,  as  illustrated in   Fig.~\ref{fig1}?  The latter is  Kramers' escape problem \cite{kramers1940brownian, hanggi1986escape}, which models, inter alia, biochemical reactions and the escape of particles from  bounded domains \cite{hanggi1990reaction, Bressloff2013,Bressloff2014}.      Although
it is well understood how long it takes for a particle to
escape a metastable state, see e.g. Refs.~\cite{Gardiner, Sidney2001, Grebenkov2014}, little is
known about the average work done on a particle when it
escapes a metastable state.

%  needs to exert on average on the particle  in order  to trigger its  escape  from the metastable state.     

%How much work must an external agent do on a mesoscopic system to let a certain event happen, such as,  the  escape of a  colloidal particle  from a metastable?    

 %triggered by a change in a potential landscape   \cite{kramers1940brownian, hanggi1986escape}, as illustrated in   Fig.~\ref{fig1}.      This question is important  in transition rate theory \cite{hanggi1990reaction}, which models   chemical reactions as a first-passage process of reactants that diffuse over a potential barrier.     While quite some research has been dedicate to  the statistics of the first-passage times \cite{kramers1940brownian, hanggi1986escape, hanggi1990reaction}, not much is known about the average work required to trigger the escape of a particle.  

Stochastic thermodynamics is a thermodynamic theory for mesoscopic systems~\cite{jarzynski1997nonequilibrium, jarzynski1997equilibrium, crooks1998nonequilibrium, crooks1999nonequilibrium, maes2003origin,  jarzynski2011equalities, Sek2010, seifert2012stochastic, van2015ensemble} and provides    experimentally testable  predictions for  their fluctuating properties \cite{Ciliberto, Gladrow}.    An important  result  in stochastic thermodynamics is the second-law-like bound \cite{jarzynski1997nonequilibrium, jarzynski1997equilibrium}
\begin{eqnarray}
\langle W(t)\rangle \geq  f\left(\lambda_{\rm f}\right) - f\left(\lambda_{\rm i}\right)\label{eq:secondx}
\end{eqnarray} 
on the  average work $\langle W(t)\rangle$ done on a system in a fixed time interval $[0,t]$ as a function of  
 the free energy difference  between the final and initial states, characterized by parameters $\lambda_{\rm f} = \lambda(t)$ and $\lambda_{\rm i} = \lambda(0)$, respectively.   In what follows we denote random variables with uppercase letters and deterministic variables with lowercase letters.      Averages $\langle \cdot\rangle$ are over repeated realizations of the process.

 Unfortunately, the bound  given by Eq.~(\ref{eq:secondx})  does not provide much insight on the average  work    $\langle W(T) \rangle = \langle \int^T_0\dot{W}(t){\rm d}t\rangle$   done on the system at an event of interest.   Indeed,  the time  $T$  when an event --- such as the escape of a particle from a metastable state --- takes place will be different for each realization of the process, and therefore the second law given by Eq.~(\ref{eq:secondx}) does not apply.    

 In this paper we    derive a  fundamental bound  on the average work an external agent has done on a system at   times $T$ when an event  happens, which we call a {\it stopping time}.     This law reads 
          \begin{eqnarray}
            \langle W(T) \rangle  \geq  \langle  f(\lambda(T))\rangle - f(\lambda_{\rm i}) +  \beta^{-1} \langle  \pi(T)\rangle, \label{eq:main}
\end{eqnarray} 
where $\langle  \pi(T)\rangle$ is a correction term that accounts for the fact that the process is in general out of equilibrium  at the stopping time $T$, and whose precise form we will specify later.        We call this law the  {\it second law of thermodynamics at stopping times}.   
 To derive the second law given by Eq.~(\ref{eq:main}),   we develop   a   thermodynamic theory for events  in nonstationary processes that take place at random times and which relies on martingale theory \cite{Chetrite2011, neri2017statistics, neri2019integral}.

 %Since 
  
\paragraph{System setup.}
 We consider a  mesoscopic system composed of slow and fast degrees of freedom.   
 The fast, internal  degrees of freedom  are hidden, whereas the   slow degrees of freedom are observed and take values in  $\mathcal{X}$.     
 
  We assume that the system  interacts weakly with an environment that is  in a state of thermal equilibrium at temperature $1/\beta$.     For a given value of the external parameter $\lambda$, the system admits an equilibrium state 
 \begin{eqnarray}
 p_{\rm eq}(x;\lambda) =  e^{-\beta [\phi(x;\lambda)-f(\lambda)]}, \quad x\in\mathcal{X}, \label{eq:Boltz}
  \end{eqnarray}  
    where $\phi$ is the free energy for a fixed value $x$  of the slow degrees of freedom and where $f$ is the free energy of the total system.    The free energy   
 \begin{eqnarray}
\phi(x;\lambda) =  u(x;\lambda) -  s_{\rm int}(x;\lambda)/\beta
 \end{eqnarray}  
 is the sum of the internal energy $u$ and the entropy  $s_{\rm int}$ associated with the internal degrees of freedom.
 
 We assume that the system is in thermal equilibrium with its environment at $t\leq 0$, and at  time $t=0$ the system  engages with an    external protocol  that drives it out of equilibrium.    The protocol consists  in a change of the external parameter  $\lambda(t)$, such that,  $\lambda(t) = \lambda_{\rm i}$ for $t\leq 0$ and $\lambda(t) = \lambda_{\rm f}$ for $t\geq \tau$.  
 
 We assume that the   internal degrees of freedom equilibrate on time scales that are much shorter than those over which $\lambda(t)$  varies (the protocol is quasi-static with respect to the internal degrees of freedom). 
 
    We aim to quantify the work done on the system at the moment when a certain event  happens (for example the escape of a particle from a metastable state).   The time when an event   happens is modelled with  a stopping time $T$.       We say that a random time $T\in [0,\infty) \cup \left\{+\infty\right\}$  is a stopping time if it is a  deterministic function defined on the set of  trajectories  $X^{+\infty}_{0} = \left\{X(t)\right\}_{t\in \mathbb{R}^+}$  that obeys causality; in other words, the value of the stopping time $T$ is independent of the outcomes of the process  $X$ after the stopping time.   If the event does not occur, then $T=+\infty$~\cite{Williams1991, Protter,  Liptser2013}. 
    
 The probability measure $\mathbb{P}$  describes the probability of events in the forward dynamics (i.e., with the protocol $\lambda(t)$ and initial distribution $p_{\rm eq}$) and we denote expectation values with respect to this measure by $\langle \cdot \rangle_{\mathbb{P}}  = \langle \cdot \rangle$.
     
    %  The system disengages from the protocol at the stopping time $T$.  

      % at the end tiem $\tau$ or when the stoppig itme.  
    % However, the   dynamics of the process after the event   $T$  is in gneral of littlre  relevance to us: once the particle escapes the metastable state it will explore a new region of phase space with a negligable probability fo returning to the original metastable. tstate.    
     %After the system has disengaged from the protocol, it relaxes to the equilibrium state 
 % \begin{eqnarray}
% p_{\rm eq}(x;\lambda_T) =  e^{-\beta [\phi(x;\lambda(T))-f(T)]}, \quad x\in\mathcal{X},
 %\end{eqnarray}
% with free energy $f(T)$.         
%    The function  $\phi(x;\lambda)$ is the  free energy of  the system when  the slow degrees of freedom are frozen in a fixed configuration $X(t) = x$.     
%Since $\phi$ is a free energy,  it is  a sum of   energy  and   entropy, 
 % \begin{eqnarray}
  %\phi(x;\lambda(t)) =  u(x;\lambda(t)) - \beta^{-1} s_{\rm int}(x;\lambda(t)).
% \end{eqnarray} 
%  If there are no internal degrees of freedom, then  $\phi(x;\lambda(t))=  u(x;\lambda(t))$.    
   
   \paragraph{Time-reversibility and martingales.}
An important feature of mesoscopic systems  is that they are  {\it time-reversible}.    Time-reversibility is defined relative to the {\it  backward} dynamics that we define    
 as follows \cite{seifert2012stochastic}:  the  state is  in the equilibrium state $p_{\rm eq}(x;\lambda_{\rm f})$ for all times $t<0$ and is subsequently driven out of equilibrium by the protocol $\tilde{\lambda}(t) = \lambda(\tau-t)$.

 The dynamics of a mesoscopic system is  time reversible if   
 there exists a  process  $S(t)$, defined on the set of trajectories $X^t_0$, such that 
       \begin{eqnarray}
  \langle A(t) \rangle_{\mathbb{P}} = \langle A(t)e^{S(t)}\rangle_{\tilde{\mathbb{P}}\circ\Theta}   \label{eq:timer}
\end{eqnarray}  
holds for any observable $A(t)$  that is a function of $X^t_0$, where the measure $\tilde{\mathbb{P}}$ describes the statistics of the process in the backward dynamics.     The map $\Theta$ is the time-reversal map that mirrors trajectories relative to the time point $\tau/2$, such that $\Theta\left[ X^{+\infty}_{-\infty}\right] =  \left\{X(\tau-t)\right\}_{t\in \mathbb{R}}$.    In other words, the expectation value of an observable in the forward dynamics can be expressed in terms of the expectation value of the same observable in the backward dynamics, as long as it is   properly reweighted with the process  $e^{S(t)}$. 

The Eq.~(\ref{eq:timer})  implies that
       \begin{eqnarray}
e^{-S(t)}  =  \Big\langle \frac{\tilde{p}
\left[\Theta(X^{+\infty}_{-\infty})\right]}{p\left[X^{+\infty}_{-\infty}\right]} \Big| X^t_0   \Big \rangle_{\mathbb{P}} \label{eq:condS}
\end{eqnarray}     
where $\tilde{p}
\left[\Theta(X^{+\infty}_{-\infty})\right]/p\left[X^{+\infty}_{-\infty}\right]$ is  the   Radon-Nikodym derivative between the two measures $\tilde{\mathbb{P}}\circ \Theta$ and $\mathbb{P}$ \cite{Liptser2013}, or loosely said, the ratio between the two associated probability densities, and where  $\Big\langle \cdot  \Big| X^t_0   \Big \rangle_{\mathbb{P}}$ is a  conditional expectation   given $X^t_0 $.   The quantity $e^{-S(t)} $  exists as long as the two measures   $\tilde{\mathbb{P}}\circ \Theta$  and  $\mathbb{P}$ are mutually absolutely continuous, which holds since the interval $[0,\tau]$ is finite and the microscopic laws of physics are time reversible.

Equation (\ref{eq:condS}) implies that  $e^{-S(t)}$ is a {\it regular martingale}.  Martingales are stochastic processes that model a gambler's fortune in a fair game of chance \cite{snell1982gambling} or    stock prices in efficient capital markets~\cite{leroy1989efficient}.     We say that a stochastic process $M(t)$ is a martingale relative to another  stochastic process $X(t)$ if: (i)~the process $M(t)$ is a real-valued  function on the set of trajectories $X^t_0$; (ii)~the process $M(t)$ is integrable, i.e., $\langle |M(t)|\rangle <\infty$; (iii)~the process $M(t)$ has no drift, i.e., with probability one $\langle M(t)|X^s_0\rangle = M(s)$ for all $s<t$  \cite{Doob1953, Doob1971, Williams1991, Liptser2013}.

  An important class of martingales are regular martingales \cite{Doob1940, Liptser2013}.   Let $Y$  be an integrable, real-valued random variable that is a function of the trajectory $X^{+\infty}_{-\infty}$.   Then the process 
  \begin{eqnarray}
  M(t) = \langle Y| X^t_0\rangle,  \quad t\in \mathbb{R}^+ ,\label{eq:regular}
  \end{eqnarray}
is a regular martingale, where $\langle \cdot | \cdot\rangle$ denotes a conditional expectation.    The martingality of $\langle Y| X^t_0\rangle$ is a direct consequence of the tower property   of conditional expectations, viz., $\langle \langle Y| X^t_0\rangle |X^s_0 \rangle = \langle Y| X^s_0\rangle $ for all $s\leq t$.

   \paragraph{Doob's optional stopping theorem and a second-law like relation at stopping times.}
 A useful property of regular martingales is Doob's optional stopping theorem, which states that for a regular martingale $M(t)$ and for a stopping time $T$ it holds that  $\langle M(T)\rangle = \langle M(0)\rangle$, see Theorem  3.2 in  Ref.~\cite{Liptser2013}.    Doob's optional stopping theorem implies that a gambler cannot make a fortune by quitting a fair game of chance at an intelligently chosen moment $T$.

      Applying Doob's optional stopping theorem to $e^{-S(t)}$,  we obtain the following integral fluctuation relation at stopping times,  
       \begin{eqnarray}
\langle e^{-S(T)}\rangle = \langle e^{-S(0)} \rangle = 1.\label{eq:integral}
\end{eqnarray}
   Using Eq.~(\ref{eq:integral}) and Jensen's inequality $\langle e^{-S(T)}\rangle \geq e^{-\langle S(T)\rangle}$, we obtain 
         \begin{eqnarray}
         \langle S(T)\rangle \geq 0,     \label{eq:second}
\end{eqnarray}
which is a second-law-like inequality.

   \paragraph{Principle of local detailed balance.} 
   The Eq.~(\ref{eq:second})  is similar to a  second law of thermodynamics, but misses a  connection with the work done on the system.    We use the  principle of local detailed balance \cite{crooks1998nonequilibrium, crooks1999nonequilibrium, maes2003origin,  jarzynski2011equalities, Sek2010, seifert2012stochastic, van2015ensemble} to link $S(t)$ with the work $W(t)$.   We say that a process obeys local detailed  balance if $S(t)$ is the total entropy production, i.e., 
   \begin{eqnarray}
   S(t) &=& -\beta Q(t) + s_{\rm int}(X(t);\lambda(t)) - s_{\rm int}(X(0);\lambda_{\rm i}) 
   \nonumber\\ 
    && 
    - \log \tilde{p}_{\tau-t}(X(t)) + \log p_{\rm eq}(X(0); \lambda_{\rm i}).  
   \end{eqnarray}
   The first term on the right-hand side is the  dissipated heat divided by the temperature and equals the change in the  environment entropy.   The second and third terms denote the change in the internal entropy (associated with the  internal degrees of freedom) and the last two terms are the change in system  entropy  (associated with the observed degrees of freedom).     The distribution $\tilde{p}_{\tau-t}(x)$ is the probability distribution of the time-reversed process at time $\tau-t$ (with external parameter $\tilde{\lambda}(\tau-t)$).    If $t\geq\tau$, then $\tilde{p}_{\tau-t}(x) = p_{\rm eq}(x;\lambda_{\rm f})$, whereas if $t<\tau$ then  $\tilde{p}_{\tau-t}(x)$ is obtained by evolving the state $p_{\rm eq}(x;\lambda_{\rm f})$ over a time interval $s\in [0,\tau-t]$ using the time-reversed protocol $\tilde{\lambda}(s) = \lambda(\tau-s)$.  
Using the first law of thermodynamics 
\begin{eqnarray}
 Q(t) +   W(t) =   u(X(t);\lambda(t)) -   u(X(0);\lambda_{\rm i}) 
  \end{eqnarray}
and the Boltzmann distribution, given by Eq.~(\ref{eq:Boltz}), we obtain  the expression (see Supplemental Material \cite{Supplement})
\begin{eqnarray}
S(t) = \beta[W(t) -   f(\lambda(t))  + f(\lambda_{\rm i})]  - \pi(t) \label{eq:SW}
\end{eqnarray}
where 
\begin{eqnarray}
 \pi(t) = \log \frac{\tilde{p}_{\tau-t}(X(t))}{p_{\rm eq}(X(t);\lambda(t))}.
\end{eqnarray}

  \paragraph{Second law of thermodynamics at stopping times.}
The Eq.~(\ref{eq:second}) together with Eq.~(\ref{eq:SW})
implies the  second law of thermodynamics at stopping times Eq.~(\ref{eq:main}) where 
 \begin{eqnarray}
 \langle  f(\lambda(T))\rangle =   \int^{\infty}_{0}{\rm d}t\: p_T(t)  f(\lambda(t))
 \end{eqnarray}
 and   
  \begin{eqnarray}
 \langle  \pi(T)\rangle 
 &=&   \int^{\infty}_{0} {\rm d}t \:\int_{\mathcal{X}}{\rm d}x\: p_{T,X(T)}(t, x)  \log \frac{\tilde{p}_{\tau-t}(x)}{p_{\rm eq}\left(x;\lambda(t)\right) } ,\nonumber\\ 
  \end{eqnarray} 
  is a  correction term that  accounts for the fact that at the stopping time the state may be far from thermal equilibrium.    
   The   distribution  $ p_{T,X(T)}(t, x) $ is the joint  probability distribution of $T$ and $X(T)$ in the forward dynamics and  $p_T(t)$ is the probability distribution of the stopping time $T$.

  The second law of thermodynamics at stopping times, given by Eq.~(\ref{eq:main}), is the main result of this Letter.   It  bounds  the average work that a mesoscopic system requires  to execute a certain task, which is completed at a stopping time~$T$.      It is reminiscent of second-law-like relations derived in  Ref.~\cite{neri2019integral}.  However, the paper \cite{neri2019integral} deals with stationary systems, whereas the Eq.~(\ref{eq:main}) holds for nonstationary systems.     
  
   The Eq.~(\ref{eq:integral}) together with Eq.~(\ref{eq:SW}) implies
    \begin{eqnarray}
  \langle e^{-\beta [W(T) -  f(\lambda(T)) + f(\lambda_{\rm i})] +    \pi(T) } \rangle = 1,  \label{eq:JarzStop}
  \end{eqnarray} 
which is a Jarzynski-like relation \cite{  
jarzynski1997nonequilibrium, 
jarzynski1997equilibrium} that holds at stopping times.  
  
   \paragraph{Limiting cases.}      
   In experiments or numerical simulations it can  be a daunting task to evaluate the quantity $\pi(t)$.  Fortunately, it turns out that $\pi(T) = 0$ in several limiting cases.    In these cases we obtain the appealing bound
          \begin{eqnarray}
\langle W(T) \rangle  \geq  \langle  f(\lambda(T))\rangle - f(\lambda_{\rm i}). \label{eq:second2}
     \end{eqnarray} 
     Examples of limiting cases for which Eq.~(\ref{eq:second2}) holds are when: (i)  the stopping time $T$ is larger than $\tau$.  Indeed, if $t\geq \tau$ then $\tilde{p}_{\tau-t}(x) = p_{\rm eq}(x;\lambda_{\rm f})$ and $\pi(t) = 0$;  (ii)   the driving $\lambda(t)$ is quasi-static.    In this case, $\tilde{p}_{\tau-t}(x) = p_{\rm eq}(x;\lambda(t))$ for all $t$,  such that $\pi(t)  = 0$;  (iii) the protocol is quenched (i.e., $\lambda(t) = \lambda_{\rm f}$ for  $t>0$) and the probability that $T=0$ is equal to zero (see supplemental material for a proof  \cite{Supplement}).   
     
     Interestingly, if the probability that $T=0$ is equal to zero, then $\pi(T) = 0$ for a protocol $\lambda(t)$ that changes slowly (quasi-static)    and also for a protocol $\lambda(t)$ that changes quickly (quenched).  Hence, we may  expect that $\pi(T) \approx 0$   holds  for intermediate driving speeds  too.   This can be verified through the Jarzynski relation at stopping times Eq.~(\ref{eq:JarzStop}), which simplifies into 
                 \begin{eqnarray}
  \langle e^{-\beta [W(T) -  f(\lambda(T)) + f(\lambda_{\rm i})]  } \rangle = 1   \label{eq:Jarz2}
  \end{eqnarray} 
 when $\pi(T) = 0$.
        
  In the next paragraphs, we use the second-law relations at stopping times Eqs.~(\ref{eq:main}) and (\ref{eq:second2}) to bound the work required to stretch a polymer or to let a particle escape.

     \paragraph{Stretching a polymer.}    
     We ask how much work is required to stretch a polymer to a certain length $\ell$, as is illustrated in Fig.~\ref{fig1}(a), and we apply the bound Eq.~(\ref{eq:main}) to this example.     We consider a setup where one end  of the polymer is anchored at  position $x=0$ to a substrate,  whereas the other  end is  fluctuating and described by a stochastic process $X(t)\in\mathbb{R}$.  The dangling end  of the polymer is connected with a spring  to an external agent, say a molecular motor, centered at~$\lambda(t)$.       
At   $t=0$,  the molecular motor starts to move and stretches the polymer until it reaches a length $\ell$, at which point the motor  stops moving and the second end point of the polymer  is anchored   to   the substrate.

     We assume that the dynamics of $X(t)$ is well described by a one-dimensional overdamped  Langevin equation
\begin{eqnarray}
\frac{{\rm d}X}{{\rm d}t} = - \mu\: \partial_x\phi(X;\lambda(t))    + \sqrt{2d}\:\xi(t),\quad t\geq 0, \label{eq:pol}
\end{eqnarray}    
where $\mu$ is the mobility coefficient, $d = \mu/\beta$ is the diffusion coefficient, $\xi(t)$ is a Gaussian white noise with $\langle \xi(t)\rangle = 0$ and $\langle \xi(t)\xi(t')\rangle = \delta(t-t')$, and where 
\begin{eqnarray}
\phi(x;\lambda(t))  =   \frac{\kappa_{\rm p}}{2} \: x^2 + \frac{\kappa_{\rm m}}{2} \left(x-\lambda(t)\right)^2
\end{eqnarray}
is the sum of the   free energy $ \kappa_{\rm p} x^2/2$,  of a polymer with one  of its end points anchored  to the substrate at $x=0$, and the free energy 
$\kappa_{\rm m} \left(x-\lambda(t)\right)^2/2$, of the spring that connects the dangling end point of the polymer to the molecular motor located at $\lambda(t)$.    Furthermore, we assume that at the initial time $t=0$  this polymer  system is in thermal equilibrium with its surroundings and that   the dynamics of the motor is 
given by 
\begin{eqnarray}
\lambda(t) = \lambda_{\rm i}  + (\lambda_{\rm f}-\lambda_{\rm i})\frac{1-e^{-t/\tau_{\rm prot}}}{1-e^{-\tau/\tau_{\rm prot}}}, \quad t\in[0,\tau],  \label{eq:prot}
\end{eqnarray}   
where $\tau_{\rm prot}>0$ characterises the speed of the protocol.    
The quantity    $\tau_{\rm rel} = 1/(\mu(\kappa_{\rm m}+\kappa_{\rm p}))$ is the polymer relaxation time.     If $\tau_{\rm prot} \ll \tau_{\rm rel}$, then the molecular motor quenches the polymer, whereas if $\tau_{\rm prot} \gg \tau_{\rm rel}$, then the motor stretches the polymer in a quasi-static manner.

The work the motor performs on the  polymer  is  \cite{Sek1998}
\begin{eqnarray}
W(t)= \int^t_0{\rm d}s\:\partial_\lambda \phi(X(s);\lambda(s))\: \dot{\lambda}(s). \label{eq:work}
\end{eqnarray}  

\begin{figure}[t]
\centering
{\includegraphics[width=0.25\textwidth]{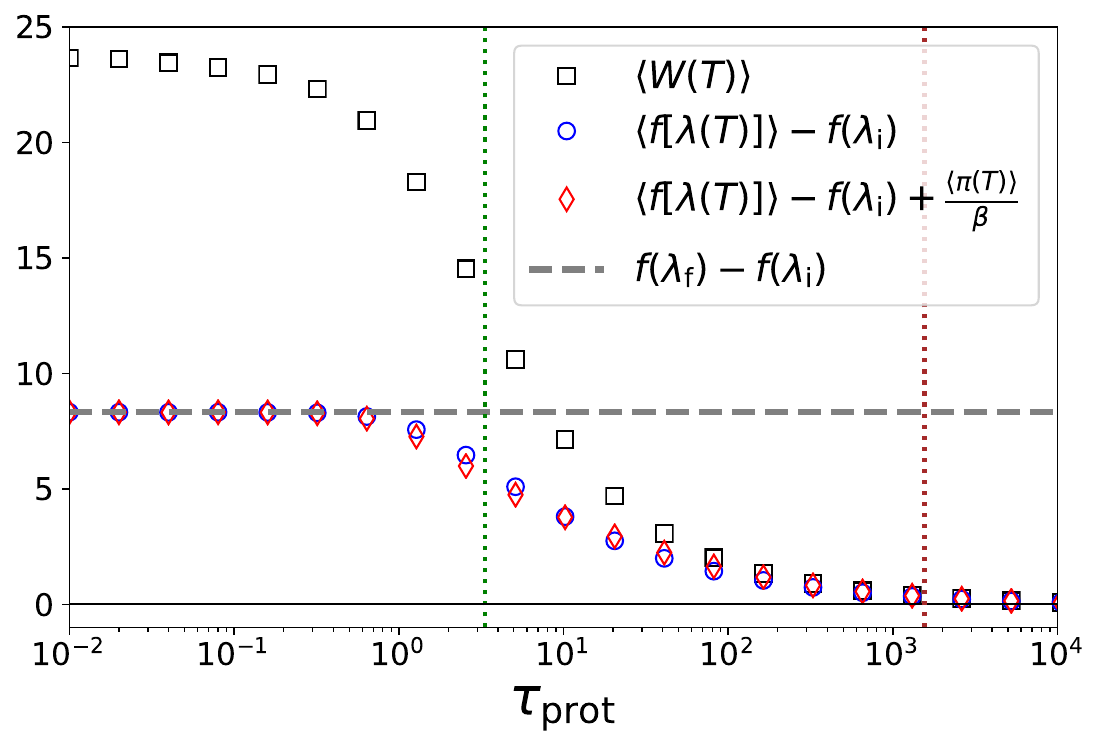}}
\put(-18,72){(a)}
{\includegraphics[width=0.25\textwidth]{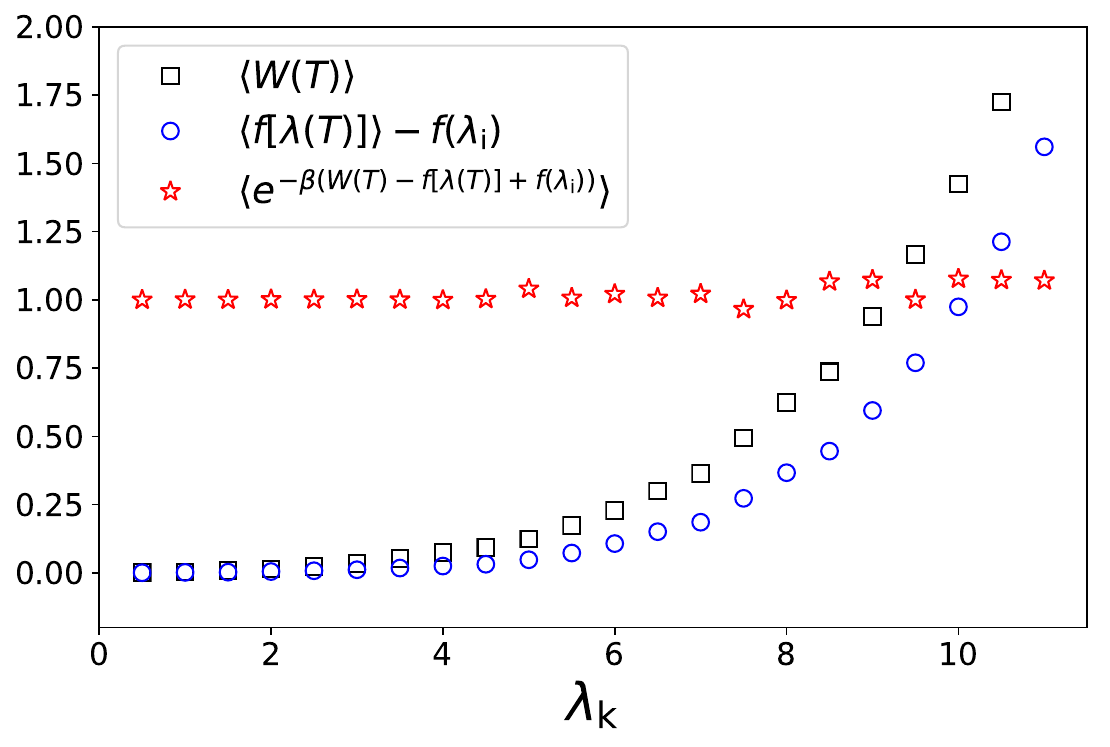}}
\put(-18,72){(b)}
\caption{Simulation results    
for  stretching a polymer (Panel (a)) and the escape problem (Panel (b)).  Panel (a): model parameters are  $\ell = 2.2$,  $\mu = 0.1$, $\beta = \kappa_{\rm p} = 1$, $\kappa_{\rm m} = 2$,   $\lambda_{\rm i} = 0.2$, $\lambda_{\rm f}= 5$, and $\tau = 1e+6$.          The relaxation time $\tau_{\rm rel} =  10/3$ and the mean first-passage time $\tau_{\rm fp} \approx 1560$ are denoted by the vertical dotted lines.    The black solid line  equals  zero and is a guide to the eye.  Panel (b):  model parameters are  $\mu = 0.1$, $\beta = x_{\rm max} = 1$, $\phi_{\rm max} = 10$, and $\tau_{\rm prot} = 4$.   Markers are averages over $1e+4$ realizations of the process.}  \label{fig2}
\end{figure}

Fig.~\ref{fig2}(a)  presents the average work $\langle W(T)\rangle$ for $T = \inf\left\{t\in[0,\tau]: |X(t)| \geq \ell  \right\}$, in other words,  the motor  stops as soon as  the polymer's length exceeds $\ell$,     
  and we compare it with the second-law-like bound Eq.~(\ref{eq:main}) (see Supplemental Material for details  \cite{Supplement}).  Interestingly, we  observe that for all values of $\tau_{\rm prot}$ the term $\langle   \pi(T)\rangle\approx 0$ and  that $\langle W(T)\rangle \geq \langle  f(\lambda(T))\rangle - f(\lambda_{\rm i})$, consistent  with the bound~Eq.~(\ref{eq:second2}).    As discussed in the previous paragraph, this can be understood from the fact that  if $\mathbb{P}(T=0)=0$, then  $\pi(T)=0$ in   both the  quasi-static and quenched limits.      
  
  For $\tau_{\rm prot}$ large enough,   $\langle W(T)\rangle\rightarrow 0$.     Indeed,  if  $\tau_{\rm prot}>\tau_{\rm fp}$ --- where  $\tau_{\rm fp} =  \frac{\sqrt{\pi}\ell^2}{4d}  \frac{e^{\alpha}}{\alpha^{3/2}}$ is  the mean first-passage time $ \langle T \rangle$ when  $\lambda_{\rm f} = \lambda_{\rm i}$, with $\alpha =\beta \frac{(\kappa_{\rm p}+\kappa_{\rm m}) \ell^2}{2}$ \cite{Grebenkov2014} --- then the polymer extends  spontaneously due to thermal fluctuations and  $\langle W(T)\rangle \approx 0$.    
  
            \paragraph{Escape problem.}    
                  We determine how much work is required   to let a colloidal particle escape a metastable state, as is illustrated in Fig.~\ref{fig1}(b).        We consider a particle described by the overdamped Langevin Eq.~(\ref{eq:pol}) with potential
                  \begin{eqnarray}
\phi(x;\lambda) = (\phi_{\rm max}-\lambda)  \frac{x^2}{x^2_{\rm max}} + \lambda, \quad x\in [0,x_{\rm max}],
\end{eqnarray}          
and reflecting boundary condition at $x=0$.   
  Initially, the particle is trapped  in the metastable state with Boltzmann distribution,  given by Eq.~(\ref{eq:Boltz}), and with  $\lambda = \lambda_{\rm i} = 0$. 

We  compute  the       average work done on the particle, given by Eq.~(\ref{eq:work}), at the escape time 
$T = \inf\left\{t\geq 0 :X(t)\geq  x_{\rm max}\right\}$.
In the absence of a driving force, the particle  escapes in a time $\langle T \rangle  = \tau_{\rm fp}\sim e^{\beta \phi_{\rm max}}$, which is very large when  $\beta \phi_{\rm max}\gg 1$.  Therefore, we facilitate the particle's escape  
 with  a kick that  deforms the potential landscape as  $\lambda(t) =  \lambda_{\rm k} e^{-t/\tau_{\rm prot}} $ for $t\geq 0$.
   Interestingly,  Fig.~\ref{fig2}(b)  shows that  the  bound Eq.~(\ref{eq:second2}) is satisfied, which indicates that again $\pi(T)\approx 0$.    This is confirmed with an evaluation of the   Jarzynski Eq.~(\ref{eq:Jarz2}) at stopping times.

\paragraph{Discussion.} In mesoscopic systems, 
physical  events  often happen at random times, such as, the escape of a colloidal particle from a metastable state~\cite{Sidney2001, Bressloff2013, Bressloff2014, Grebenkov2014}.    
We have derived the  second law of thermodynamics at stopping times Eq.~(\ref{eq:main}), which  bounds  the average amount of work  that has been done on a system at a  stopping time or first-passage time~$T$ as a result of a change in the free-energy landscape.      This second law applies to arbitrary systems that
obey local detailed balance and arbitrary stopping times.

If $\langle  \pi(T)\rangle\approx 0$, then the second law     Eq.~(\ref{eq:main}) simplifies into  Eq.~(\ref{eq:second2}).  Interestingly, we have shown that  Eq.~(\ref{eq:second2}) holds in the quasi-static limit and for quenched protocols when $T>0$ with probability one.   Additionally, using numerical simulations we  find that  in our examples   Eq.~(\ref{eq:second2}) holds at  intermediate driving speeds of the protocol, and I believe this will be in general the case (as long as $T>0$ with probability one).

   If  $\langle \pi(T)\rangle< \beta [f(\lambda_{\rm i}) - \langle  f(\lambda(T))\rangle]$, then  the system can perform work on its environment.    For instance, we can  stop the process as  soon as $W(t)>\epsilon$, with $\epsilon$ a small positive number (see Supplemental Material for an example). Work extraction by stopping a process at an intelligently chosen moment 
 is closely related to the construction of  Maxwell demons, which are  smart devices that change the protocol of a system at a cleverly chosen moment \cite{Par}.        However, in the thermodynamics at stopping times   we do not consider what happens after the stopping time (e.g.~in the escape problem we are not interested in the events that happen after the  particle has escaped the potential).    

The present Letter demonstrates how  for nonstationary processes thermodynamic  relations  at stopping times can be derived  using the martingale $e^{-S(t)}$ given by Eq.~(\ref{eq:condS}); so far, thermodynamic properties of  stochastic processes at first-passage times have mainly been studied  in the context of  stationary processes \cite{roldan2015, neri2017statistics, garrahan, Gingrich, quantum, neri2019integral}.   
     It would be interesting to use  the martingality of  $e^{-S(t)}$ to   derive bounds on, e.g., extreme values of $Q(t)$  \cite{Singy} or  mean first-passage times \cite{roldan2015}  in nonstationary processes. 
\acknowledgements
The author acknowledges  A.~Barato, K.~Brander, R.~Ch\'{e}trite,  G.~Falasco, E.~Fodor, J.~Garrahan, S.~Gupta, F.~J\"{u}licher,  G.~Manzano, P.~Pietzonka, S.~Pigolotti, E.~Rold\'{a}n,  S.~Samu and S.~Singh for interesting discussions.

\clearpage
\setboolean{@twoside}{false}


\section*{Supplementary material}

# Supplemental Material for
# "Second Law of Thermodynamics at Stopping Times"

Izaak Neri

*Department of Mathematics, King's College London, Strand, London, WC2R 2LS, UK*

In a first section we derive the equation (12) that relates the entropy production $S(t)$ to the work $W(t)$. In a second section, we show that for quenched systems $\langle \pi(T) \rangle / \beta = f(\lambda_f) - \langle f[\lambda(T)] \rangle + \langle W(T) \rangle - \langle W(\tau) \rangle$, and hence for quenched systems the second law at stopping times (2) is implied by the second law at fixed times (1). Moreover, we show that if $\mathbb{P}(T = 0)$, then $\pi(T) = 0$ for quenched protocols. In a third section we detail how we compute the bound Eq. (2) for the example of the stretched polymer. Finally, in a fourth section we show how the second law (2) bounds the amount of work a system can do on its environment by stopping at a cleverly chosen moment.

### S1.  DERIVATION OF THE EXPRESSION (12) FOR $S(t)$

Our starting point is the relation (10) for $S(t)$. Using the first law of thermodynamics (11) in (10), we obtain

$$S(t) = \beta W(t) - \beta \left\{ u[X(t); \lambda(t)] - u[X(0); \lambda_i] \right\}$$
$$+ s_{\text{int}}[X(t); \lambda(t)] - s_{\text{int}}[X(0); \lambda_i] - \log \tilde{p}_{\tau-t}(x) + \log p_{\text{eq}}[X(0); \lambda_i]. \tag{S1}$$

Using the Boltzmann distribution (3) and the expression (4) for the free energy $\phi$, we obtain

$$S(t) = \beta \left\{ W(t) - f[\lambda(t)] + f(\lambda_i) \right\} - \pi(t) \tag{S2}$$

with $\pi(t)$ given by Eq. (13).

## S2. QUENCHED SYSTEMS

Consider a quenched system with the protocol

$$\lambda(t) = \begin{cases} \lambda_i, & t \leq 0, \\ \lambda_f, & t > 0. \end{cases} \tag{S3}$$

### A. Generic expression for $\langle \pi(T) \rangle / \beta$

We show that

$$\langle \pi(T) \rangle / \beta = f(\lambda_f) - \langle f[\lambda(T)] \rangle + \langle W(T) \rangle - \langle W(\tau) \rangle, \tag{S4}$$

where $\tau$ is now an arbitrary positive time. As a consequence, the second law at stopping times (2) is equivalent to the second law at fixed times (1).

Let $\mathcal{X}_0 \in \mathcal{X}$ be the region of phase space for which $T = 0$. We denote the probability that $T > 0$ by

$$\gamma := \mathbb{P}(T > 0) = \int_{x \in \mathcal{X} \setminus \mathcal{X}_0} dx\, p_{\mathrm{eq}}(x; \lambda_i). \tag{S5}$$

Since the protocol only changes at $t = 0$, we can simplify the quantities that appear in the second-law-like bound (2). The average free energy at the stopping time equals

$$\langle f[\lambda(T)] \rangle = \gamma f(\lambda_f) + (1 - \gamma) f(\lambda_i). \tag{S6}$$

and the average work at the stopping time

$$\langle W(T) \rangle = \int_{x \in \mathcal{X} \setminus \Delta_0} \mathrm{d}x\, p_{\mathrm{eq}}(x; \lambda_i) \left[ \phi(x; \lambda_f) - \phi(x; \lambda_i) \right]. \tag{S7}$$

Note that if the system has internal degrees of freedom, then $\beta \langle Q(T) \rangle = \langle s_{\mathrm{int}}[X(T); \lambda(T)] \rangle - \langle s_{\mathrm{int}}[X(0); \lambda_i] \rangle$: although the protocol (S3) is implemented instantaneously with regard to the slow degrees of freedom, we assume that the protocol is quasistatic with regard to the fast internal degrees of freedom.

Let's now consider the quantity of interest $\langle \pi(T) \rangle$, which depends on $\tilde{p}_{\tau-t}(x)$. Since we quench the system, $\tilde{p}_{\tau-t}(x) = p_{\mathrm{eq}}(x; \lambda_f)$ for all $t \geq 0$, and therefore $\pi(t) = 0$ for all $t > 0$. As a consequence

$$\langle \pi(T) \rangle = \int_{x \in \mathcal{X}_0} \mathrm{d}x\, p_{\mathrm{eq}}(x; \lambda_i) \log \frac{p_{\mathrm{eq}}(x; \lambda_f)}{p_{\mathrm{eq}}(x; \lambda_i)}. \tag{S8}$$

Moreover, since

$$p_{\text{eq}}(x; \lambda_i) = e^{-\beta[\phi(x;\lambda_i)-f(\lambda_i)]}, \quad p_{\text{eq}}(x; \lambda_f) = e^{-\beta[\phi(x;\lambda_f)-f(\lambda_f)]}, \tag{S9}$$

we obtain that

$$\begin{aligned}\langle \pi(T) \rangle / \beta &= (1-\gamma)[f(\lambda_f) - f(\lambda_i)] \\ &\quad - \int_{x \in \mathcal{X}_0} \mathrm{d}x \, p_{\text{eq}}(x; \lambda_i) \left[\phi(x; \lambda_f) - \phi(x; \lambda_i)\right].\end{aligned} \tag{S10}$$

Using (S6) we identify

$$(1-\gamma)[f(\lambda_f) - f(\lambda_i)] = f(\lambda_f) - \langle f[\lambda(T)] \rangle \tag{S11}$$

and using (S7) we obtain

$$\begin{aligned}\langle \pi(T) \rangle / \beta &= f(\lambda_f) - \langle f[\lambda(T)] \rangle + \langle W(T) \rangle \\ &\quad - \int_{x \in \mathcal{X}} \mathrm{d}x \, p_{\text{eq}}(x; \lambda_i) \left[\phi(x; \lambda_f) - \phi(x; \lambda_i)\right].\end{aligned} \tag{S12}$$

Moreover, since

$$\langle W(\tau) \rangle = \int_{x \in \mathcal{X}} \mathrm{d}x \, p_{\text{eq}}(x; \lambda_i) \left[\phi(x; \lambda_f) - \phi(x; \lambda_i)\right] \tag{S13}$$

we obtain the equality (S4), which is what we aimed to prove.

### B.  $\langle \pi(T) \rangle = 0$ when $\mathbb{P}[T=0] = 0$

Since $\gamma = 1 - \mathbb{P}[T=0] = 1$ we obtain from (S6) that

$$\langle f(\lambda(T)) \rangle = f(\lambda_f). \tag{S14}$$

Moreover, since $\mathbb{P}[T=0] = \mathbb{P}[\mathcal{X}_0] = \int_{x \in \mathcal{X}_0} \mathrm{d}x \, p_{\text{eq}}(x; \lambda_i) = 0$, we obtain from (S7) and (S13) that

$$\langle W(T) \rangle = \langle W(\tau) \rangle. \tag{S15}$$

Finally, using (S14) and (S15) in (S4) we obtain that $\langle \pi(T) \rangle = 0$, which is what we were meant to prove.

## S3. STRETCHED POLYMER: SECOND LAW OF THERMODYNAMICS AT STOPPING TIMES

We detail how we compute the right-hand side of the second law, Eq. (2), for the model (19) with free energy (20).

We show that the total free energy difference

$$\langle f[\lambda(T)]\rangle - f(\lambda_i) = \frac{\kappa_m \kappa_p}{\kappa_m + \kappa_p}\left[\langle \lambda^2(T)\rangle - \lambda_i^2\right], \tag{S16}$$

and that

$$\begin{aligned}\langle \pi(T)\rangle/\beta &= -\frac{(\lambda_f - \lambda_i)^2}{2}\left(\frac{\tau_{\rm rel}}{\tau_{\rm prot} + \tau_{\rm rel}}\right)^2 \frac{\kappa_m^2}{\kappa_p + \kappa_m}\left[\int_0^\tau dt\, p_T(t) e^{-2t/\tau_{\rm prot}}\right] \\ &+ \kappa_m(\lambda_f - \lambda_i)\frac{\tau_{\rm rel}}{\tau_{\rm prot} + \tau_{\rm rel}}\left[\int_0^\tau dt\, p_{T,X(T)}(t,x)\left(x - \langle x\rangle_{{\rm eq},\lambda(t)}\right) e^{-t/\tau_{\rm prot}}\right] \\ &+ O(e^{-\tau/\tau_{\rm prot}}).\end{aligned} \tag{S17}$$

In the first subsection we derive the formula (S16). In the second and third subsections we derive explicit expressions for the probability density functions in the forward and backward dynamics by solving the Smoluchowski equation. Finally, in the fourth subsection, we derive the formula (S17).

### 1. Equilibrium properties

The equilibrium state of the polymer and spring and described with the free energy (17) is the Gaussian distribution

$$p_{\rm eq}(x;\lambda) = \frac{1}{\sqrt{2\pi\sigma_\lambda^2}} e^{-\frac{(x - \langle x\rangle_{{\rm eq},\lambda})^2}{2\sigma^2}}, \quad x \in \mathbb{R}, \tag{S18}$$

with mean value

$$\langle x\rangle_{{\rm eq},\lambda} = \int_\mathbb{R} dx\, x\, p_{\rm eq}(x;\lambda) = \frac{\kappa_m \lambda}{\kappa_p + \kappa_m} \tag{S19}$$

and variance

$$\sigma^2 = \int_\mathbb{R} dx\, x^2 p_{\rm eq}(x;\lambda) - \langle x\rangle^2 = \frac{1}{\beta(\kappa_p + \kappa_m)}. \tag{S20}$$

Note that the variance is independent of $\lambda$.

The equilibrium free energy

$$\beta f(\lambda(t)) = \frac{1}{2}\log\frac{\kappa_{\mathrm{m}}+\kappa_{\mathrm{p}}}{2\pi\mathsf{T}_{\mathrm{env}}} + \frac{1}{2}\frac{\kappa_{\mathrm{m}}\kappa_{\mathrm{p}}}{\kappa_{\mathrm{m}}+\kappa_{\mathrm{p}}}\lambda^2(t), \tag{S21}$$

and the free energy difference

$$\beta f[\lambda(t)] - \beta f(\lambda_i) = \frac{1}{2}\frac{\kappa_{\mathrm{m}}\kappa_{\mathrm{p}}}{\kappa_{\mathrm{m}}+\kappa_{\mathrm{p}}}\left[\lambda^2(t) - \lambda^2(0)\right]. \tag{S22}$$

### 2. Probability density of $X(t)$ in the forward dynamics

The probability density in the original, forward dynamics is described by the Smoluchowski equation

$$\partial_t p_t(x) = \mu(\kappa_{\mathrm{p}}+\kappa_m)p_t(x) + \mu[\kappa_{\mathrm{p}}x + \kappa_m(x - \lambda(t))]\partial_x p_t(x) + d\,\partial_x^2 p_t(x) \tag{S23}$$

with initial condition $p_0(x) = p_{\mathrm{eq}}(x;\lambda_i)$, $t \geq 0$, and $x \in \mathbb{R}$. This equation is solved by

$$p_t(x) = \frac{1}{\sqrt{2\pi\sigma^2}}e^{-\frac{(x-m(t))^2}{2\sigma^2}}, \tag{S24}$$

where $\sigma^2$ is given by (S20) and where

$$m(t) = \langle x \rangle_{\mathrm{eq},\lambda_i}\,e^{-t/\tau_{\mathrm{rel}}} + \frac{1}{\tau_{\mathrm{rel}}}\frac{\kappa_{\mathrm{m}}}{\kappa_{\mathrm{p}}+\kappa_{\mathrm{m}}}\int_0^t e^{(s-t)/\tau_{\mathrm{rel}}}\lambda(s)\mathrm{d}s \tag{S25}$$

with $\tau_{\mathrm{rel}} = 1/[\mu(\kappa_{\mathrm{p}}+\kappa_{\mathrm{m}})]$. For the protocol (18) we obtain for $t \in [0,\tau]$ that

$$\begin{aligned}
m(t) &= \frac{\kappa_m}{\kappa_p+\kappa_m}\lambda_i\left[e^{-t/\tau_{\mathrm{rel}}} + \frac{e^{-t/\tau_{\mathrm{rel}}}-1}{e^{\tau/\tau_{\mathrm{prot}}}-1}\right] \\
&+ \frac{\kappa_m}{\kappa_p+\kappa_m}\frac{e^{\tau/\tau_{\mathrm{prot}}}}{e^{\tau/\tau_{\mathrm{prot}}}-1}\left[\lambda_f + \frac{\lambda_f\tau_{\mathrm{rel}} - \lambda_i\tau_{\mathrm{prot}}}{\tau_{\mathrm{prot}}-\tau_{\mathrm{rel}}}e^{-t/\tau_{\mathrm{rel}}}\right. \\
&\left. + e^{-t/\tau_{\mathrm{prot}}}\frac{\tau_{\mathrm{prot}}}{\tau_{\mathrm{prot}}-\tau_{\mathrm{rel}}}(\lambda_i-\lambda_f)\right].
\end{aligned} \tag{S26}$$

Note that in the quasi-static limit with $\tau/\tau_{\mathrm{prot}} \to \infty$ and $\tau_{\mathrm{prot}}/\tau_{\mathrm{rel}} \to \infty$, it holds that $m(t) = \langle x \rangle_{\mathrm{eq},\lambda(t)}$.

In figure S1 we illustrate the dynamical properties of the stretched polymer for parameters similar to those of Fig.1(a). In particular, we compare the mean position $\langle x \rangle_{\mathrm{eq},\lambda(t)}$ in the equilibrium state $p_{\mathrm{eq}}(x;\lambda(t))$ with the mean position $m(t)$ in the nonequilibrium state $p(x;\lambda(t))$. We observer that $\langle x \rangle_{\mathrm{eq},\lambda(t)} \approx m(t)$ in the quasi-static regime but that $\langle x \rangle_{\mathrm{eq},t}$ and $m(t)$ can significantly deviate when the protocol is implemented faster.

<=""></=""><=""><=""></=""></=""><=""><=""><=""></=""></=""></="">

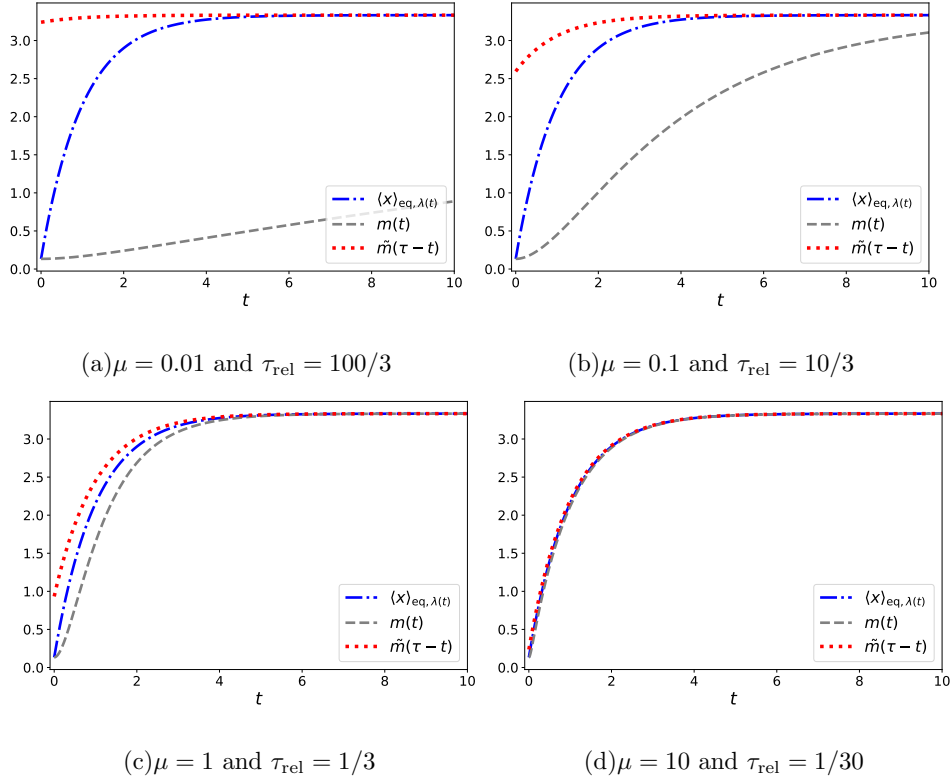

FIG. S1: Stretching a polymer: illustration of $\langle x \rangle_{\text{eq},t}$, $m_t$ and $\tilde{m}_t$ for the parameters $\beta = \kappa_{\text{p}} = 1$, $\kappa_{\text{m}} = 2$, $\lambda_i = 0.2$, $\lambda_f = 5$, $\tau_{\text{prot}} = 1$, $\tau = 20$ and for $\mu$ as given. The subfigure (a) is in the quenched limit of $\tau_{\text{rel}}/\tau_{\text{prot}} \gg 1$ whereas the subfigure (b) concerns the quasi-static limit of $\tau_{\text{prot}}/\tau_{\text{rel}} \gg 1$.

### 3. Probability density of $X(t)$ in the backward dynamics

The probability density in the conjugate, time-reversed dynamics is described by the Smoluchowski equation

$$\partial_t \tilde{p}_t = \mu(\kappa_{\text{p}} + \kappa_m)\tilde{p}_t(x) + \mu(\kappa_{\text{p}} x + \kappa_m(x - \lambda(\tau - t)))\partial_x \tilde{p}_t(x) + d\,\partial_x^2 \tilde{p}_t(x), \quad \text{(S27)}$$

with initial condition $\tilde{p}_0(x) = p_{\text{eq}}(x; \lambda_f)$, $t \geq 0$, and $x \in \mathbb{R}$. Therefore,

$$\tilde{p}(x;t) = \frac{1}{\sqrt{2\pi\sigma^2}} e^{-\frac{(x-\tilde{m}(t))^2}{2\sigma^2}}, \quad \text{(S28)}$$

where

$$\tilde{m}(t) = \langle x \rangle_{\text{eq},\lambda_f} e^{-t/\tau_{\text{rel}}} + \frac{1}{\tau_{\text{rel}}} \frac{\kappa_{\text{m}}}{\kappa_{\text{p}} + \kappa_{\text{m}}} \int_0^t e^{(s-t)/\tau_{\text{rel}}} \lambda(\tau - s)\mathrm{d}s. \quad \text{(S29)}$$

For the protocol (18) we obtain for times $t \in [0, \tau]$ that

$$\tilde{m}(t) = \frac{\kappa_m}{\kappa_p + \kappa_m}\lambda_f$$
$$+ \frac{\kappa_m}{\kappa_p + \kappa_m}\frac{\lambda_f - \lambda_i}{e^{\tau/\tau_{\text{prot}}} - 1}\left[1 - \frac{\tau_{\text{rel}}}{\tau_{\text{prot}} + \tau_{\text{rel}}}e^{-t/\tau_{\text{rel}}} - \frac{\tau_{\text{prot}}}{\tau_{\text{prot}} + \tau_{\text{rel}}}e^{t/\tau_{\text{prot}}}\right] \quad \text{(S30)}$$

and thus

$$\tilde{m}(\tau - t) = \frac{\kappa_m}{\kappa_p + \kappa_m}\lambda_f$$
$$+ \frac{\kappa_m}{\kappa_p + \kappa_m}\frac{\lambda_f - \lambda_i}{e^{\tau/\tau_{\text{prot}}} - 1}\left[1 - \frac{\tau_{\text{rel}}}{\tau_{\text{prot}} + \tau_{\text{rel}}}e^{-(\tau-t)/\tau_{\text{rel}}} - \frac{\tau_{\text{prot}}}{\tau_{\text{prot}} + \tau_{\text{rel}}}e^{(\tau-t)/\tau_{\text{prot}}}\right]. \quad \text{(S31)}$$

In the limit of $\tau \gg \tau_{\text{prot}}$, we obtain

$$\tilde{m}(\tau - t) = \frac{\kappa_m}{\kappa_p + \kappa_m}\left[\lambda_f + (\lambda_i - \lambda_f)\frac{\tau_{\text{prot}}}{\tau_{\text{prot}} + \tau_{\text{rel}}}e^{-t/\tau_{\text{prot}}}\right] + O(e^{-\tau/\tau_{\text{prot}}})$$
$$= \langle x \rangle_{\text{eq},\lambda(t)} + \frac{\kappa_m}{\kappa_p + \kappa_m}(\lambda_f - \lambda_i)\frac{\tau_{\text{rel}}}{\tau_{\text{prot}} + \tau_{\text{rel}}}e^{-t/\tau_{\text{prot}}} + O(e^{-\tau/\tau_{\text{prot}}}). \quad \text{(S32)}$$

In the regime of a quasi-static driving ($\tau_{\text{rel}} \ll \tau_{\text{prot}}$) it holds that $\tilde{m}(\tau - t) = \langle x \rangle_{\text{eq},\lambda(t)} + O(\tau_{\text{rel}}/\tau_{\text{prot}})$.

In figure S1 we compare the mean value $\tilde{m}(\tau - t)$ of the polymer length in the conjugate dynamics with those in the forward dynamics [$m(t)$] and in the equilibrium state [$\langle x \rangle_{\text{eq},\lambda(t)}$]. We observe that for quasistatic driving $\tilde{m}(\tau - t) = m(t) = \langle x \rangle_{\text{eq},\lambda(t)}$, but otherwise the three values $\langle x \rangle_{\text{eq},\lambda(t)}$, $m(t)$ and $\tilde{m}(\tau - t)$ can be significantly different.

### 4. The quantity $\langle \pi(T) \rangle$

The relations (15), (S18), (S19), (S20), (S28), and (S29) imply that

$$\langle \pi(T) \rangle/\beta = \left\langle \frac{(X(T) - \langle x \rangle_{\text{eq},\lambda(T)})^2}{2\sigma^2} \right\rangle - \left\langle \frac{(X(T) - \tilde{m}(\tau - T))^2}{2\sigma^2} \right\rangle$$
$$= \frac{\kappa_p + \kappa_m}{2}\left[\int_0^\tau dt\, p_T(t)(\langle x \rangle^2_{\text{eq},\lambda(t)} - \tilde{m}^2(\tau - t))\right]$$
$$+ (\kappa_p + \kappa_m)\left[\int_0^\tau dt\, p_{T,X(T)}(t, x)\, x(\tilde{m}(\tau - t) - \langle x \rangle_{\text{eq},\lambda(t)})\right]. \quad \text{(S33)}$$

If $\tau \gg \tau_{\rm prot}$, then we can use (S32) and thus

$$
\begin{aligned}
\langle \pi(T) \rangle / \beta &= -\frac{(\lambda_f - \lambda_i)^2}{2} \left( \frac{\tau_{\rm rel}}{\tau_{\rm prot} + \tau_{\rm rel}} \right)^2 \frac{\kappa_m^2}{\kappa_p + \kappa_m} \left[ \int_0^\tau dt\, p_T(t) e^{-2t/\tau_{\rm prot}} \right] \\
&+ \kappa_m (\lambda_f - \lambda_i) \frac{\tau_{\rm rel}}{\tau_{\rm prot} + \tau_{\rm rel}} \left[ \int_0^\tau dt\, p_{T,X(T)}(t,x) \left( x - \langle x \rangle_{\rm eq, \lambda(t)} \right) e^{-t/\tau_{\rm prot}} \right] \\
&+ O(e^{-\tau/\tau_{\rm prot}}).
\end{aligned}
\tag{S34}
$$

In the example of Fig. 1(a) with the stopping time $T = \inf\{t > 0 : |X(t)| \geq \ell\}$, we have that $p_{T,X(T)}(t,x) = p_T(t)\delta(x;\ell)$ and the distribution $p_T(t)$ fully determines $\langle \pi(T) \rangle / \beta$.

## S4. EXTRACTING WORK FROM A MESOSCOPIC SYSTEM BY STOPPING AT A CLEVERLY CHOSEN MOMENT

Using numerical simulations we illustrate how an experiment can be constructed for which $\langle W(T) \rangle < 0 < \langle \Delta f(T) \rangle$, such that the mesoscopic system exerts force on its surroundings while the average free energy difference $\langle \Delta f(T) \rangle$ is positive. We consider the example of a stretched polymer described by model (19) with free energy (20).

Fig. S2 shows that work extraction from stretching a polymer is possible when using a stopping time

$$T = \inf\{t > 0 : X(t) \leq \ell\}. \tag{S35}$$

In this case, an average negative work is reached when the speed of the protocol is slower than the polymer relaxation time [$\tau_{\rm prot} > \tau_{\rm rel}$], as demonstrated in Fig. S2 [the average extracted work $\langle W(T) \rangle$ is denoted by the black squares]. Moreover, the results in Fig. S2 show that it is possible to extract work from a mesoscopic system even when the free energy difference is positive by stopping at a cleverly chosen moment (the blue circles denote the average free energy difference). However, one cannot extract more work than determined by the second law of thermodynamics at stopping times given by Eq. (2), which is shown by the red diamonds in Fig. S2.

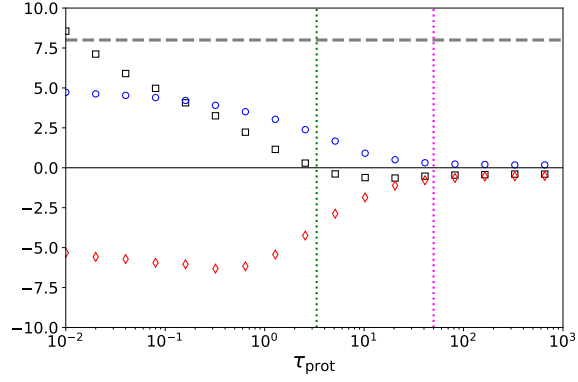

FIG. S2: Illustration of work extraction in a stretched polymer experiment modelled by the Langevin equation (19) with free energy (20). Simulation results shown are for the average work $\langle W(T) \rangle$ (black squares), the average free energy difference $\langle f[\lambda(T)] \rangle - f(\lambda_i)$ (blue circles), the right-hand-side $\langle f[\lambda(T)] \rangle - f(\lambda_i) + \langle \pi(T) \rangle / \beta$ of the second law (2) (red diamonds), and the total free energy difference $f(\lambda_f) - f(\lambda_i)$ (dashed grey line) as a function of the speed of the protocol $\tau_{\text{prot}}$. The solid black line is a guide to the eye that denotes the value 0. The stopping time used is $T = \min\{t > 0 : X(t) \leq \ell\}$. Parameters are $\ell = 0.2$, $\mu = 0.1$, $\beta = 0.1$, $\kappa_p = 1$, $\kappa_m = 2$, $\lambda_i = 1$, $\lambda_f = 5$, $\tau = 50$. The values of $\tau_{\text{rel}} = 10/3$ and $\tau = 50$ are indicated by the vertical dotted lines.




\begin{thebibliography}{0}
\expandafter\ifx\csname natexlab\endcsname\relax\def\natexlab#1{#1}\fi
\expandafter\ifx\csname bibnamefont\endcsname\relax
  \def\bibnamefont#1{#1}\fi
\expandafter\ifx\csname bibfnamefont\endcsname\relax
  \def\bibfnamefont#1{#1}\fi
\expandafter\ifx\csname citenamefont\endcsname\relax
  \def\citenamefont#1{#1}\fi
\expandafter\ifx\csname url\endcsname\relax
  \def\url#1{\texttt{#1}}\fi
\expandafter\ifx\csname urlprefix\endcsname\relax\def\urlprefix{URL }\fi
\providecommand{\bibinfo}[2]{#2}
\providecommand{\eprint}[2][]{\url{#2}}

\end{thebibliography}


\begin{thebibliography}{10}

\bibitem{kramers1940brownian}
H.~A. Kramers, {\it Brownian motion in a field of force and the diffusion model of
  chemical reactions}, Physica {\bf 7}, 284-304 (1940).

\bibitem{hanggi1986escape}
P.~H{\"a}nggi, {\it  Escape from a metastable state},  Journal of Statistical
  Physics {\bf 42}, 105-148 (1986).
  
  
      
\bibitem{hanggi1990reaction}
P.~H{\"a}nggi, P.~Talkner and M.~Borkovec,
\newblock \emph{Reaction-rate theory: fifty years after Kramers},
\newblock Reviews of modern physics \textbf{62}(2), 251 (1990).
  
  \bibitem{Bressloff2013}
  P.C.~Bressloff, and J.M.~Newby,  {\it  Stochastic models of intracellular transport}, Reviews of Modern Physics {\bf 85}, 135  (2013). 
  
    \bibitem{Bressloff2014}
      P.C.~Bressloff,  {\it Stochastic Processes in Cell Biology} (Springer, Berlin, 2014).
      
  

\bibitem{Gardiner}C.~Gardiner, {\it Stochastic Methods} (Springer, Berlin, 2009).


  \bibitem{Sidney2001}S.~Redner, {\it  A Guide to First-Passage Processes} (Cambridge University Press, Cambridge, 2001).

  
\bibitem{Grebenkov2014}D.~S.~Grebenkov, {\it First exit times of harmonically trapped particles: a didactic review}, Journal of Physics A: Mathematical and Theoretical {\bf 48},  013001 (2015).


  
\bibitem{jarzynski1997nonequilibrium}
C.~Jarzynski, {\it Nonequilibrium equality for free energy differences}, 
  Physical Review Letters {\bf 78}, 2690 (1997).
  


\bibitem{jarzynski1997equilibrium}
C.~Jarzynski, {\it Equilibrium free-energy differences from nonequilibrium
  measurements: A master-equation approach},  Physical Review E {\bf 56},
5018 (1997).



\bibitem{crooks1998nonequilibrium}
G.~E.~Crooks, {\it Nonequilibrium measurements of free energy differences for
  microscopically reversible markovian systems},  Journal of Statistical
  Physics {\bf 90},  1481-1487 (1998).
  
  \bibitem{crooks1999nonequilibrium}G.~E.~Crooks, {\it Entropy production fluctuation theorem and the nonequilibrium work relation for free energy differences},  Physical Review E  {\bf 60},  2721 (1999).


  \bibitem{maes2003origin}
C.~Maes, {\it On the origin and the use of fluctuation relations for the
  entropy}, S{\'e}minaire Poincar{\'e} {\bf 2}, 29-62 (2003).
  
    \bibitem{Sek2010}K.~Sekimoto, {\it Stochastic Energetics} (Springer, Berlin, Germany, 2010).


\bibitem{jarzynski2011equalities}
C.~Jarzynski, {\it Equalities and inequalities: irreversibility and the second law
  of thermodynamics at the nanoscale}, Annual Review of Condensed Matter Physics~{\bf 2}, 329-351 (2011).

\bibitem{seifert2012stochastic}
U.~Seifert, {\it Stochastic thermodynamics, fluctuation theorems and molecular
  machines},  Reports on Progress in Physics {\bf 75},  126001
  (2012).
  

\bibitem{van2015ensemble}
C.~Van~den Broeck and M.~Esposito, {\it Ensemble and trajectory thermodynamics: A
  brief introduction}, Physica A: Statistical Mechanics and its
  Applications {\bf 418}, 6-16 (2015).
  
  
\bibitem{Ciliberto}S.~Ciliberto, {\it Experiments in stochastic thermodynamics: Short history and perspectives},  Physical Review X {\bf 7}, 021051  (2017).

  
 \bibitem{Gladrow} J. Gladrow, M. Ribezzi-Crivellari, F. Ritort, and U. F.
Keyser, {\it Experimental evidence of symmetry breaking of
transition-path times}, Nat. Commun. {\bf 10}, 55 (2019).

\bibitem{Chetrite2011}R.~Ch\'{e}trite and S.~Gupta, {\it 
 Two refreshing views of fluctuation theorems through kinematics elements and exponential martingale},  Journal of Statistical Physics {\bf 143}, 543 (2011).
 

\bibitem{neri2017statistics}
I.~Neri, E.~Rold{\'a}n, and F.~J{\"u}licher, {\it Statistics of infima and
  stopping times of entropy production and applications to active molecular
  processes}, Physical Review X {\bf 7}, 011019 (2017).
  


\bibitem{neri2019integral}
I.~Neri, \'{E}.~Rold{\'a}n, S.~Pigolotti, and F.~J{\"u}licher, {\it Integral
  fluctuation relations for entropy production at stopping times},  Journal of Statistical Mechanics: Theory and Experiment {\bf 2019}, 104006 (2019). 

  

\bibitem{Doob1953}
J.~L.~Doob, {\it Stochastic Processes},  (John Wiley \& Sons, Chapman \& Hall, New York,  USA, 1953)


  \bibitem{Williams1991}
 D.~Williams, {\it Probability with Martingales},   (Cambridge University Press, Cambridge, UK, 1991).

\bibitem{Liptser2013}
R.~Liptser and A.~N.~Shiryaev, {\it Statistics of Random Processes: I. General Theory}, 2nd ed. (Springer Science \& Business Media, Berlin, 2013), Vol. {\bf 5}.

\bibitem{Protter}P.~E.~Protter, {\it Stochastic Integration and Differential Equations},  (Springer, Berlin, 2005).



  
  


\bibitem{snell1982gambling}
J.~L.~Snell, {\it Gambling, Probability and Martingales},  The Mathematical
  Intelligencer {\bf 4}, 118-124 (1982).

\bibitem{leroy1989efficient}
S.~F.~LeRoy, {\it Efficient capital markets and martingales},  Journal of
  Economic Literature {\bf 27}, 1583-1621 (1989).
 

\bibitem{Doob1971}
J.~L.~Doob, {\it What is a Martingale?},  The American Mathematical Monthly {\bf 78}, 451-463 (1971). 




\bibitem{Doob1940}J.~L.~Doob, {\it Regularity properties of certain families of chance variables}, Transactions of the American Mathematical Society {\bf 47},  455-486 (1940).


\bibitem{Supplement}{See Supplemental Material for
a derivation of the second law of thermodynamics at stopping times, a discussion of this law for quenched systems, a discussion of the application of this law for the  stretched polymer, and  an illustration of work extraction by stopping
a stochastic process at a cleverly chosen moment.}



  
\bibitem{Sek1998}K.~Sekimoto, {\it Langevin equation and thermodynamics}, Progress  of Theoretical  Physics Supplements {\bf 130}, 17 (1998).

   
   
   
\bibitem{Par}J.~M.~R.~Parrondo, J.~M.~Horowitz, and T.~Sagawa, {\it Thermodynamics of information}, Nature physics {\bf 11},  131-139 (2015).

\bibitem{garrahan}J.~P.~Garrahan, {\it Simple bounds on fluctuations and uncertainty relations for first-passage times of counting observables}, Physical Review E {\bf 95}, 032134 (2017).

\bibitem{Gingrich}T.~R.~Gingrich, and J.~M.~Horowitz, {\it Fundamental bounds on first passage time fluctuations for currents}, Physical Review Letters {\bf 119}, 170601 (2017). 

\bibitem{quantum}G.~Manzano, R.~Fazio, and E.~Rold\'{a}n, {\it Quantum martingale theory and entropy production},   Physical Review Letters {\bf 122}, 220602 (2019).

\bibitem{Singy}Singh, Shilpi, et al., {\it Extreme reductions of entropy in an electronic double dot},  Physical Review B {\bf 99}, 115422 (2019).

   \bibitem{roldan2015}E.~Rold\'{a}n, I.~Neri, M.~D\"{o}rpinghaus, H.~Meyr, and F.~J\"{u}licher, {\it Decision making in the arrow of time},  Physical Review Letters {\bf 115},  250602 (2015).
   

     
\end{thebibliography}
 \end{document}